\documentclass[conference]{IEEEtran}
\IEEEoverridecommandlockouts
\usepackage{cite}
\usepackage{amsmath,amssymb,amsfonts}
\usepackage{mathtools}
\usepackage{algorithmic}
\usepackage{graphicx}
\usepackage{textcomp}
\usepackage{xcolor}
\usepackage{hyperref}
\usepackage{braket}
\usepackage{amsthm}
\usepackage{tikz}
\usepackage{tikz-3dplot}

\providecommand{\myvec}[1]{\ensuremath{\boldsymbol{#1}}}

\providecommand{\tt}{\ensuremath{\myvec{t}}}

\newtheorem{theorem}{Theorem}

\newtheorem{definition}[theorem]{Definition}

\def\BibTeX{{\rm B\kern-.05em{\sc i\kern-.025em b}\kern-.08em
    T\kern-.1667em\lower.7ex\hbox{E}\kern-.125emX}}

\renewcommand{\figureautorefname}{Figure~\negthinspace}
\renewcommand{\equationautorefname}{Equation~\negthinspace}
\renewcommand{\tableautorefname}{Table~\negthinspace}

\begin{document}

\title{Learning to Program Quantum Measurements for Machine Learning
\thanks{The views expressed in this article are those of the authors and do not represent the views of Wells Fargo. This article is for informational purposes only. Nothing contained in this article should be construed as investment advice. Wells Fargo makes no express or implied warranties and expressly disclaims all legal, tax, and accounting implications related to this article. 

This work is also supported by the U.S. DOE, under award DE-SC-0012704 and BNL’s LDRD \#24-061. This research used resources of the NERSC, under Contract No. DE-AC02-05CH11231 using NERSC award HEP-ERCAP0033786.}
}

\author{
\IEEEauthorblockN{Samuel Yen-Chi~Chen}
\IEEEauthorblockA{\textit{Wells Fargo} \\
New York NY, USA \\
yen-chi.chen@wellsfargo.com}
\and 
\IEEEauthorblockN{Huan-Hsin~Tseng}
\IEEEauthorblockA{\textit{AI Department} \\
\textit{Brookhaven National Laboratory}\\
Upton NY, USA  \\
htseng@bnl.gov}
\and
\IEEEauthorblockN{Hsin-Yi~Lin}
\IEEEauthorblockA{\textit{Department of Mathematics} \\
\textit{and Computer Science} \\
\textit{Seton Hall University}\\
South Orange NJ, USA \\
hsinyi.lin@shu.edu}
\and
\IEEEauthorblockN{Shinjae~Yoo}
\IEEEauthorblockA{\textit{AI Department} \\
\textit{Brookhaven National Laboratory}\\
Upton NY, USA \\
syjoo@bnl.gov}
}

\maketitle

\begin{abstract}
The rapid advancements in quantum computing (QC) and machine learning (ML) have sparked significant interest, driving extensive exploration of quantum machine learning (QML) algorithms to address a wide range of complex challenges. The development of high-performance QML models requires expert-level expertise, presenting a key challenge to the widespread adoption of QML. Critical obstacles include the design of effective data encoding strategies and parameterized quantum circuits, both of which are vital for the performance of QML models. Furthermore, the measurement process is often neglected-most existing QML models employ predefined measurement schemes that may not align with the specific requirements of the targeted problem. We propose an innovative framework that renders the observable of a quantum system—specifically, the Hermitian matrix-trainable. This approach employs an end-to-end differentiable learning framework, enabling simultaneous optimization of the neural network used to program the parameterized observables and the standard quantum circuit parameters. Notably, the quantum observable parameters are dynamically programmed by the neural network, allowing the observables to adapt in real time based on the input data stream. Through numerical simulations, we demonstrate that the proposed method effectively programs observables dynamically within variational quantum circuits, achieving superior results compared to existing approaches. Notably, it delivers enhanced performance metrics, such as higher classification accuracy, thereby significantly improving the overall effectiveness of QML models.
\end{abstract}


\section{Introduction}
Quantum computing (QC) promises computational capabilities that surpass classical computing for a range of intractable problems \cite{nielsen2010quantum}. In parallel, recent advances in artificial intelligence (AI) and machine learning (ML) have enabled intelligent agents to achieve remarkable performance in domains such as sequential decision-making \cite{silver2017mastering} and natural language processing \cite{vaswani2017attention}. Motivated by the complementary strengths of these two paradigms, a growing body of research has focused on hybrid quantum-classical frameworks aimed at harnessing the best of both worlds. While today’s quantum hardware has yet to realize large-scale, fault-tolerant quantum algorithms, hybrid frameworks such as variational quantum algorithms (VQAs) provide a practical bridge for near-term applications \cite{bharti2022noisy,cerezo2021variational}.

VQAs represent a hybrid computing paradigm in which computational tasks are decomposed into subcomponents delegated to either quantum or classical processors, depending on their respective strengths. Quantum subroutines are typically implemented using variational quantum circuits (VQCs), while classical routines handle gradient computation and optimization. VQAs have demonstrated effectiveness in applications such as quantum chemistry \cite{peruzzo2014variational} and combinatorial optimization \cite{farhi2014quantum}.

Building on this foundation, VQCs have been widely adopted as learnable modules in quantum machine learning (QML), enabling end-to-end training pipelines where VQC parameters are updated using classical gradient methods \cite{mitarai2018quantum,schuld2020circuit}. VQC-based QML has shown promise across diverse ML tasks, including classification \cite{mitarai2018quantum,schuld2020circuit,chen2021end,chen2022quantumCNN,qi2023qtnvqc,lin2024QuantumGradCAM}, sequence modeling and regression \cite{chen2022quantumLSTM,chen2024qeegnet}, generative modeling \cite{chu2023iqgan}, natural language processing \cite{li2023pqlm,yang2022bert,di2022dawn,stein2023applying}, model compression and quantum distillation \cite{liu2024quantum,lin2024quantum,liu2024qtrl,liu2024federated,liu2024quantum2,lin2024quantum2,liu2024quantum3,chen2024_QT_DIST_RL,liu2024_QT_FWP}, and reinforcement learning \cite{chen2022variationalQRL,chen2020QRL,chen2024efficient,meyer2022survey,chen2023quantumLSTM_RL,lockwood2020reinforcement,skolik2021quantum,jerbi2021variational,coelho2024vqc,CHEN2023321Async,yun2023quantum}.

Despite this progress, most quantum neural networks (QNNs) are still trained using fixed observables $\hat{B}_k$—typically the Pauli-$X$, $Y$, or $Z$ operators—whose spectral range is limited to $\lambda = \pm 1$. As a result, the output of a VQC is confined to $\langle \psi | H | \psi \rangle \in [-1, 1]$, regardless of the specific parameterization of the quantum circuit $U(\vec{x}), W(\Theta)$. According to the Rayleigh quotient, we know that $\lambda_{\min} \leq \langle \psi | H | \psi \rangle \leq \lambda_{\max}$ for any normalized state $|\psi\rangle$, suggesting that increasing the spectral range of the observable can expand the expressivity of the VQC—a desirable property for tasks such as classification and regression.

Prior work has demonstrated that parameterizing and training the observable-represented as a Hermitian matrix-alongside the VQC parameters can yield substantial gains over fixed measurements \cite{chen2025learning_to_measure}. However, in this setting, the observable is static during inference, which limits its adaptability. While this is more flexible than fixed Pauli observables, it still cannot adjust to data-specific measurement needs.

\begin{figure}[htbp]
\begin{center}
\includegraphics[width=1\columnwidth]{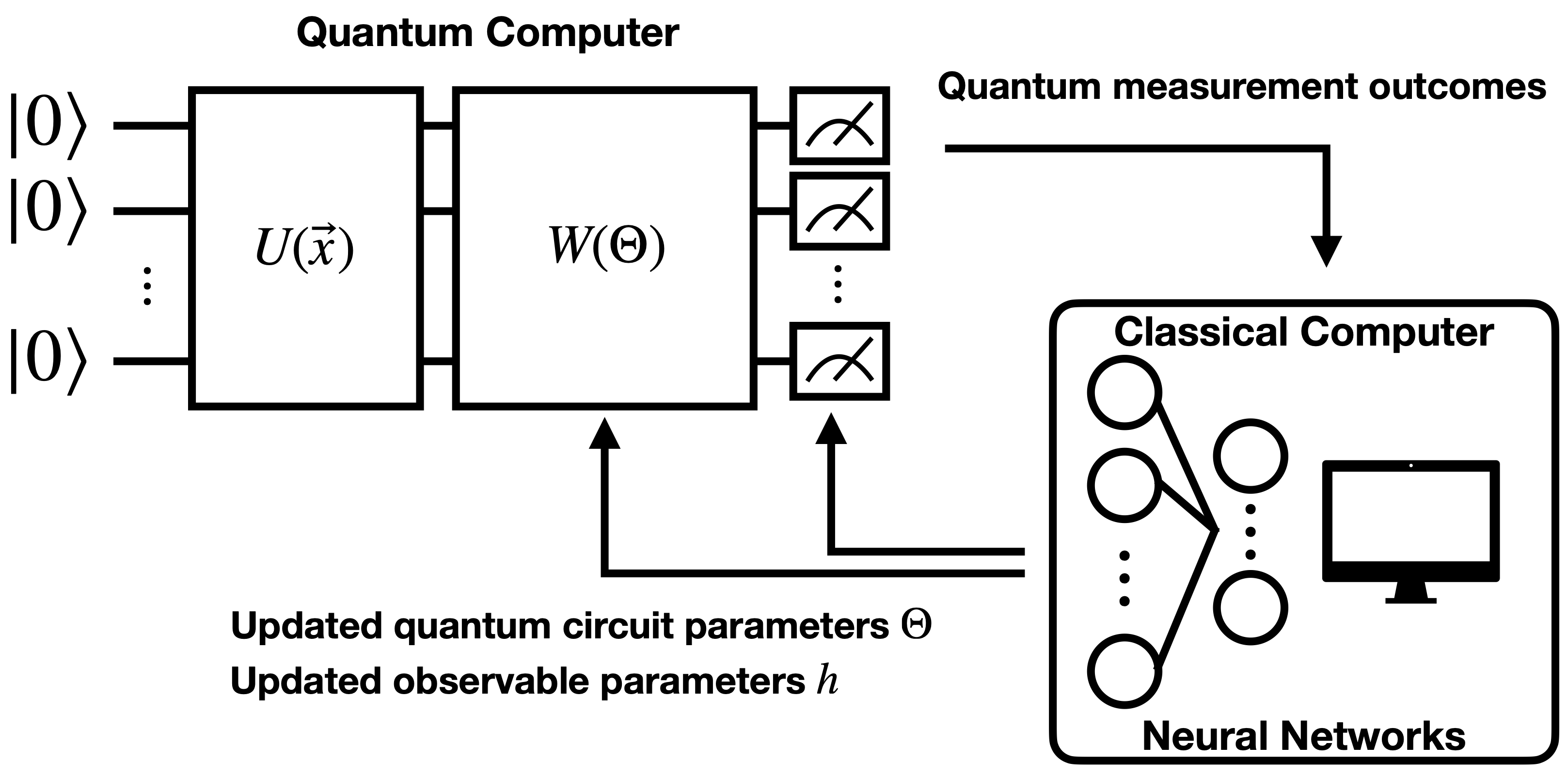}
\caption{{\bfseries Hybrid quantum-classical computing with learnable observable programmed by the neural network.}}
\label{fig:hybrid_computing}
\end{center}
\end{figure}

In this paper, we propose a dynamic framework in which an external neural controller, referred to as the slow programmer, is trained to generate both the VQC rotation parameters and the observable (Hermitian matrix) on a per-input basis, as illustrated in \figureautorefname{\ref{fig:hybrid_computing}}. This transforms the VQC into a data-reactive quantum model whose measurement scheme adapts to the input distribution. Numerical simulations across various classification benchmarks show that the proposed method consistently outperforms existing alternatives, offering both higher accuracy and improved training stability.
\section{Related Work}
Prior work has demonstrated that learning the quantum observable, represented as a parameterized Hermitian matrix, can significantly enhance the performance of variational quantum circuits (VQCs). In particular, \cite{chen2025learning_to_measure} showed that making the observable trainable leads to substantial gains over fixed choices such as the Pauli-$Z$ matrix. Moreover, using separate optimizers for circuit parameters and observable parameters further improves convergence stability and final accuracy. This idea is rooted in the broader trend of parameterizing and optimizing more components of a quantum model beyond rotation angles. One prominent direction is differentiable quantum architecture search (DiffQAS) \cite{martyniuk2024quantum,zhang2022differentiable,chen2024differentiable}, which adapts the classical differentiable architecture framework \cite{liu2018darts} to the quantum domain by assigning trainable weights to candidate operations, allowing joint structure-parameter optimization. Beyond direct gradient-based optimization, the Fast Weight Programmer (FWP) framework \cite{schmidhuber1992learning} introduces a meta-learning mechanism in which a ``slow'' neural controller dynamically generates parameters for a ``fast'' neural actor. This paradigm has been extended to quantum settings in Quantum FWP (QFWP) \cite{chen2024learning,liu2024_QT_FWP}, where a classical controller network generates data-dependent parameters for a VQC. The present work advances this line by enabling the slow neural controller to program not only the quantum circuit parameters but also the measurement observable. This transforms the entire hybrid quantum-classical model into a data-driven, reactive system. Additionally, unlike previous works where the observable is fixed after training, our framework supports on-the-fly observable adaptation at inference time. Together, these extensions enable a more expressive, flexible, and dynamically reprogrammable quantum learning architecture.
\section{Quantum Neural Networks}
A typical variational quantum circuit (VQC) comprises three primary components: an encoding circuit $U(\vec{x})$ that maps classical input data into quantum states, a parameterized variational circuit $W(\Theta)$ consisting of trainable quantum gates, and a final quantum measurement step using one or more observables $\hat{B}_k$. The encoding circuit prepares the initial state $U(\vec{x})\ket{0}^{\otimes n}$, where $\vec{x}$ is the input feature vector and $n$ is the number of qubits. This encoded state is then transformed by the variational circuit, yielding the final quantum state
\begin{equation}
\label{eqn:vqc_state_psi}
    \ket{\Psi} = W(\Theta) U(\vec{x})\ket{0}^{\otimes n},
\end{equation}
where the variational circuit is composed of $M$ layers of trainable unitary blocks:
\begin{equation}
\label{eqn:vqc_state_psi_2}
    W(\Theta)=  \prod_{j = M}^{1} W_{j}(\vec{\theta_{j}}), \quad \Theta = \{\vec{\theta_{1}} \cdots \vec{\theta_{M}}\}.
\end{equation}
Information is extracted from the quantum system via measurement of Hermitian operators $\hat{B}_k$, yielding expectation values $\langle \hat{B}_k \rangle = \bra{\Psi} \hat{B}_k \ket{\Psi}$. In practice, this expectation can be evaluated through repeated measurements on real quantum hardware (sampling over multiple shots), or computed directly in simulation. The VQC can thus be interpreted as a quantum function $\vec{f}(\vec{x}; \Theta) = (\langle \hat{B}_1 \rangle, \dots, \langle \hat{B}_K \rangle)$, where the choice of observable $\hat{B}_k$ determines how quantum information is projected back into the classical domain. As we will discuss, replacing fixed observables with trainable or input-conditioned operators opens up new expressive capabilities for quantum models.
\begin{figure}[htbp]
\begin{center}
\includegraphics[width=1\columnwidth]{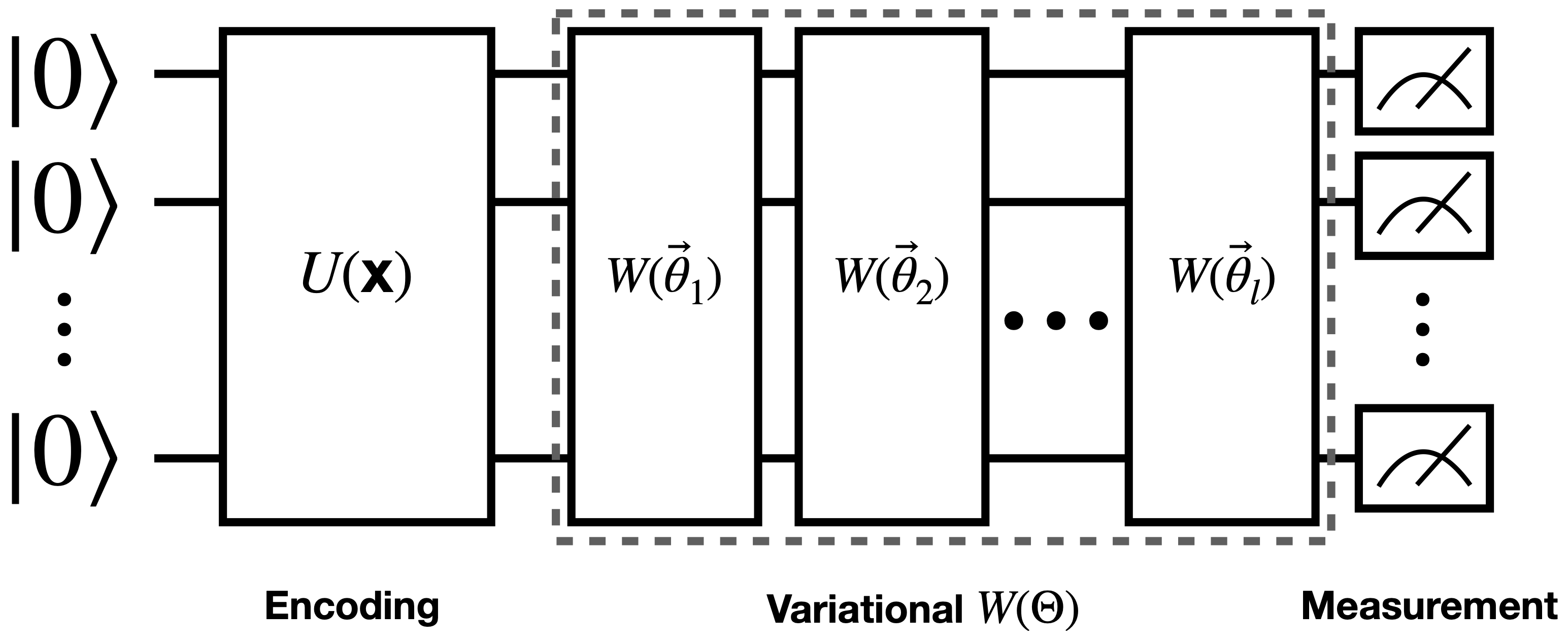}
\caption{{\bfseries Basic structure of a variational quantum circuit (VQC).}}
\label{fig:vqc_scheme}
\end{center}
\end{figure}
\section{Programmable Quantum Measurements}
A typical assumption in machine learning is that data reside on a differentiable manifold~\cite{roweis2000nonlinear, belkin2003laplacian}. When classical data are transformed to quantum states, their geometric structure are mapped to the Hilbert space of the quantum domain accordingly. To extract the intrinsic structure of data in the quantum setting, we introduce a dynamically evolving inner product defined via a fibration in an extended geometric space. This formulation considers the space of quantum states as a complex vector bundle, where the metric structure adapts in response to variations in the underlying data manifold. This perspective naturally aligns with gauge-theoretic principles~\cite{bleeckergauge, tseng2018gravitational}, wherein the choice of fiber structure and the corresponding connection provide insights into quantum embeddings and their evolution.

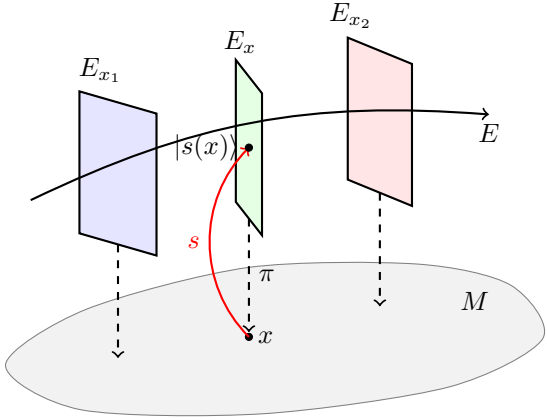
\begin{figure}[htbp]
 \vskip -0.2in
  \centering
\pgfmathsetmacro{\zplane}{-2.5}

\tdplotsetmaincoords{70}{150}
\begin{tikzpicture}[tdplot_main_coords, scale=0.67]

  \coordinate (A1) at (0.17, -0.98, 0);
  \coordinate (B1) at (-0.17, 0.98, 0);
  \coordinate (C1) at (-0.17, 0.98, 3);
  \coordinate (D1) at (0.17, -0.98, 3);
  \draw[fill=red!20, opacity=0.5] (A1) -- (B1) -- (C1) -- (D1) -- cycle;
  \draw[thick] (A1) -- (B1) -- (C1) -- (D1) -- cycle;
   \node at (0.1, -0.98, 3.) [above] {$E_{x_2}$};
  
  \coordinate (A2) at (2.74, -0.97, 0);
  \coordinate (B2) at (3.26, 0.97, 0);
  \coordinate (C2) at (3.26, 0.97, 3);
  \coordinate (D2) at (2.74, -0.97, 3);
  \draw[fill=green!20, opacity=0.5] (A2) -- (B2) -- (C2) -- (D2) -- cycle;
  \draw[thick] (A2) -- (B2) -- (C2) -- (D2) -- cycle;
  \node at (3.2, 0, 3.33) [above] {$E_x$};
  
  \coordinate (A3) at (6.34, -0.94, 0);
  \coordinate (B3) at (5.66, 0.94, 0);
  \coordinate (C3) at (5.66, 0.94, 3);
  \coordinate (D3) at (6.34, -0.94, 3);
  \draw[fill=blue!20, opacity=0.5] (A3) -- (B3) -- (C3) -- (D3) -- cycle;
  \draw[thick] (A3) -- (B3) -- (C3) -- (D3) -- cycle;
    \node at (6.4, 0, 3.3) [above] {$E_{x_1}$};
  
  \draw[fill=gray!20, opacity=0.5] plot [smooth cycle] coordinates {
    (-2,-2,\zplane)
    (1,-3.5,\zplane)
    (5,-3,\zplane)
    (8,-1,\zplane)
    (8,2,\zplane)
    (5,4,\zplane)
    (1,4.5,\zplane)
    (-2,3,\zplane)
    (-3,0,\zplane)
  };
    \node at (-2.6,0.2, \zplane - 0.2) [left] {$M$};
  
    \draw[->, thick, dashed] (0,0,0) -- (0,0,\zplane + 0.1);
      \draw[->, thick, dashed] (3,0,0) -- (3,0,\zplane + 0.1)node[midway, right] {$\pi$};
        \draw[->, thick, dashed] (6,0,0) -- (6,0,\zplane + 0.1);
  
  \filldraw[black] (3,0,\zplane) circle (2pt) node[right] {$x$};
  
  \draw[->, thick, red] (3,0,\zplane) to[bend left=45] node[midway, left] {$s$} (3,0,1.5);
  
  \filldraw[black] (3,0,1.5) circle (2pt) node[left] {$\ket{s(x)}$};
  
  \draw[->, thick] (8,0,1.3) to[bend left=15] (-2.5,0,1.2) node[below] {$E$};
  
\end{tikzpicture}
  \caption{Our Learning to Program Quantum Measurements is described by a vector bundle $(E, \pi, M)$ with $\pi: E \to M$ the bundle projection. On each classical data point $x\in M$, the living Hilbert space of quantum states is given by $E_x =\pi^{-1}(x) \cong \mathbb{C}^N$ along with a Hermitian inner product $h(x)$ on $E_x$ smoothly changing with respect to $x$. A local section $s: W \subseteq M \to E$ encodes (converts) $x\in M$ into a quantum state $\ket{s(x)} \in E_x$. The inner product $h(x)\left( \ket{s(x)}, \ket{s(x)} \right)$ induces a Hermitian operator $H(x)$ via $\bra{s(x)} H(x) \ket{s(x)}$.}
\label{fig: vector bundle}
\end{figure}

The mathematical description of our method is as follows. Consider a set of classical data $\mathcal{D} = \{ (x_i, y_i) \in \mathbb{R}^k \times \mathbb{R} ^m \}$ distributed on a (differentiable) manifold $M$. To conduct QML, the classical data are embedded into quantum states $\ket{\psi_{x_i} } $ of $n$-qubit Hilbert space $\mathbb{C}^N$ with $N = 2^n$. This ``classical-quantum'' conversion can be resulted from a (smooth) \emph{section} from $M$ to a complex vector bundle $(E, \pi)$ of rank $N$ with $\pi: E \to M$ the projection (see \figureautorefname{\ref{fig: vector bundle}})\cite{jost2008riemannian}. Denote a bundle chart of a point $x \in M$ by $(\varphi, W)$ with $x \in W \subseteq M$, then the fiber $E_x := \pi^{-1}(x)$ is isomorphic to a usual $n$-qubit space $\mathbb{C}^N$. Together with a local section $s: W \to E$, each sample $x \in M$ is mapped to a quantum state $s(x)$ (or $\ket{s(x)}$) that has its own qubit space $\{ x \} \times \mathbb{C}^N$.

{\color{red}}
To perform a quantum measurement, a bundle metric is required on $(E, \pi)$.

\begin{definition}\label{Def: bundle metric}
    Let $E \xrightarrow{\pi} M$ be a complex vector bundle and $\overline{E}$ be the complex conjugate vector bundle defined by $\alpha * v = \overline{\alpha} \cdot v$ for $\alpha \in \mathbb{C}$, $v \in \pi^{-1}(x)$, $x\in M$. Then a \textbf{Hermitian metric} is a smooth section $h: M \to E \otimes \overline{E}^*$ such that
    \begin{equation}\label{E: Hermitian metric}
    \begin{aligned}
        h(x)(v, w) &= \overline{h(x)(w, v)} \qquad (\text{for all $v, w \in E_x$})\\
        h(x)(v, v) &>0 \qquad (\text{for any $0 \neq v \in E_x$})
    \end{aligned}
    \end{equation}
\end{definition}
where $\overline{E}^*$ is the dual bundle of $\overline{E}$. Endowing a bundle metric on a complex vector bundle allows us to measure each quantum state \emph{dynamically}, depending on each classical sample $x \in M$. Indeed, given a frame field (basis) $\{e_j\}_{j=1}^N$ in $E$, from (\ref{E: Hermitian metric}) we see
\begin{equation}
    h(x)(e_j, e_k) = \overline{h(x)(e_k, e_j)} 
\end{equation}
which yields a Hermitian matrix $H_{jk} = \overline{H}_{kj}$ at each sample $x \in M$ by denoting $H_{jk} = h(x)(e_j, e_k)$. Conversely, whenever a smooth mapping $x \in M \mapsto H(x)$ with values in Hermitian matrices is given, we can define
\begin{equation}\label{E: pointwise Hermitian metric}
    \widetilde{h}(x)(v, w) := \sum_{j,k} \overline{v_j} \, w_k \cdot H_{jk}(x)
\end{equation}
Then one verifies that $\widetilde{h}$ is a Hermitian metric satisfying (\ref{E: Hermitian metric}). The scheme by (\ref{E: pointwise Hermitian metric}) is proposed as our ``\emph{Learning to Program Quantum Measurements}''.

Consequently, a quantum state $\ket{\psi_x} \in E_x$ corresponding to $x \in M$ has measurement value $h(x)\left( \psi_x, \psi_x \right)$ by (\ref{E: Hermitian metric}), or $\bra{\psi_x} H \ket{\psi_x} $ in Dirac notations.

In this study, a classical neural network is used to create a smooth map from $x \mapsto H(x)$ of a Hermitian matrix, \emph{i.e.,} an observable. However, since Hermitian matrices do not form a Lie group, the following parametrization method is constructed to ensure neural network predictions are Hermitian matrices.
\begin{figure}
\begin{center}
\includegraphics[width=1\columnwidth]{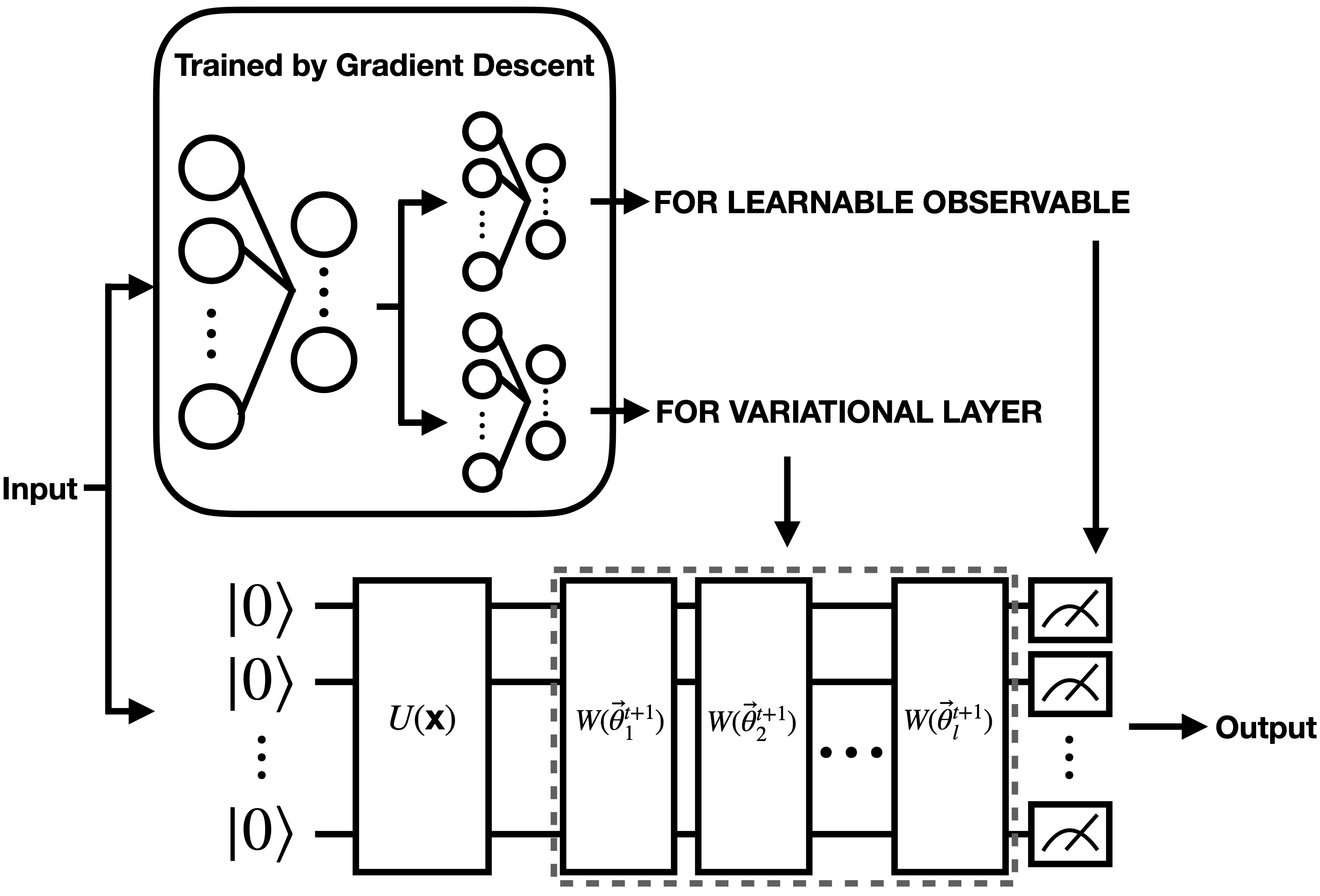}
\caption{{\bfseries Schematic of the proposed framework integrating fast weight programmers (FWP) with variational quantum circuits (VQCs).} An external neural network controller, trained via gradient descent, takes the classical input and generates two sets of parameters: one for the variational layers of the VQC and another for the learnable quantum observable. The input is encoded via $U(\vec{x})$ and processed through trainable quantum layers $W(\vec{\theta}_j)$ with dynamically generated parameters. The final measurement is performed using a data-conditioned observable, enabling fully input-adaptive quantum inference.}
\label{fig:fwp_observable}
\end{center}
\end{figure}

\begin{definition}
A Hermitian matrix $B$ is used as an \emph{observable} in quantum mechanics, for it yields real-valued expectations $\bra{\Psi} B \ket{\Psi}$ under a wave function $\ket{\Psi}$. The condition $B = B^{\dagger}$ requires each matrix element $b_{ij} = \overline{b_{ji}} \in \mathbb{C}$. Consequently, a Hermitian matrix can be generated by $2 \times \frac{N(N-1)}{2} + N = N^2$ real parameters,
\begin{equation}
    B = \begin{pmatrix}
d_{11} & a_{12} + i c_{12} & a_{13} + i c_{13} & \cdots & a_{1N} + i c_{1N}  \\
* & d_{22}  & a_{23} + i c_{23}  & \cdots & a_{2N} + i c_{2N}  \\
* & * & d_{33}  & \cdots & a_{3N} + i c_{3N}  \\
\vdots & \vdots & \vdots & \ddots & \vdots \\
* & * & * & \cdots & d_{NN}
\end{pmatrix}
\nonumber
\end{equation}
where $*$ denotes the corresponding complex conjugate and $a_{ij}$, $c_{ij}, d_{ii}$ are arbitrary real numbers.
\end{definition}
The Hermitian matrix can be initialized randomly and optimized iteratively via gradient-based methods. To elucidate the process, we can write the Hermitian matrix of a $n$-qubit system with $N = 2^n$ as the parametrization of coefficient $\vec{b} = (b_{11}, \ldots, b_{NN}) \in \mathbb{C}^{N \times N}$ as $B(\vec{b}) = \sum_{i = 1}^N\sum_{j = 1}^N b_{ij} \, E_{ij}$, where $b_{ij} = \overline{b_{ji}}$ and $E_{ij}$ as the indicating matrix with only one non-zero at entry $(i,j)$, 
\begin{equation}
E_{ij} = 
\begin{pmatrix}
0 & 0 & \cdots & 0 & 0 \\
0 & 0 & \cdots & 0 & 0 \\
\vdots  & \vdots & \ddots & \vdots & \vdots \\
0 & 0 & \cdots & 1 & 0 \\
0 & 0 & \cdots & 0 & 0
\end{pmatrix}
\end{equation}
Given a quantum state $\ket{\Psi}$ (e.g. the one shown in \equationautorefname{\ref{eqn:vqc_state_psi}}), the expectation value with the $B(\vec{b})$ can be written as,
\begin{align}
    \bra{\Psi}B(\vec{b})\ket{\Psi} 
    &= \bra{\Psi}\sum_{i}^{N} \sum_{j}^{N} b_{ij} E_{ij}\ket{\Psi} \\
    &= \sum_{i}^{N} \sum_{j}^{N} b_{ij}  \bra{\Psi}E_{ij}\ket{\Psi}
\end{align}
The total differential $\bra{\Psi}B(\vec{b})\ket{\Psi}$ is 

\begin{equation}
    \nabla \bra{\Psi}B(\vec{b})\ket{\Psi} = 
\begin{pmatrix}
    \nabla_{\theta} \bra{\Psi}B(\vec{b})\ket{\Psi}\\
    \nabla_{\vec{b}} \bra{\Psi}B(\vec{b})\ket{\Psi}
\end{pmatrix}
\end{equation}
where the differentiation of $\bra{\Psi}B(\vec{b})\ket{\Psi}$ can be written as,
\begin{align}
    &\frac{\partial \bra{\Psi} B(\vec{b}) \ket{\Psi}}{\partial b_{k\ell}}
    =\frac{\partial \bra{0}U(\vec{x})^\dagger W(\Theta)^\dagger B(\vec{h}) W(\Theta) U(\vec{x})\ket{0}}{\partial b_{k\ell}} \nonumber  \\
    &= \frac{\partial}{ \partial b_{k\ell}} \sum_{i}^{N} \sum_{j}^{N} b_{ij}  \bra{0}U(\vec{x})^\dagger W(\Theta)^\dagger E_{ij} W(\Theta) U(\vec{x})\ket{0} \nonumber \\
    &=  \sum_{i}^{N} \sum_{j}^{N} \delta_{ik}\delta_{j \ell}  \bra{0}U(\vec{x})^\dagger W(\Theta)^\dagger E_{ij} W(\Theta) U(\vec{x})\ket{0} \nonumber  \\
    &= \bra{0}U(\vec{x})^\dagger W(\Theta)^\dagger E_{k \ell }W(\Theta) U(\vec{x})\ket{0}  \nonumber \\
    &= \overline{ \big( W(\Theta) U(\vec{x}) )}_{k 1} \big( W(\Theta) U(\vec{x}) \big)_{\ell 1} \nonumber
\end{align}
where the bar is the complex conjugate and $k, \ell \in \{1, \ldots, N\}$ denote matrix indices. $\Theta$ represents the parameters of the variational quantum circuit. The last equality shows the explicit dependency on $W(\Theta)$.
As illustrated in \figureautorefname{\ref{fig:fwp_observable}}, under the FWP framework, an external neural network can be employed to generate the VQC rotation parameters, the parameters controlling the measurement observables, or both. When FWP is applied to a single component (either the rotation angles or the observable), the architecture consists of a single-layer feedforward neural network whose input size matches the feature dimension and whose output size corresponds to the number of trainable parameters for that component. In the case where FWP is used to jointly program both the VQC and the observable, the framework adopts an encoder-decoder design: the input is first encoded into a latent representation, which is then processed by two separate neural networks to produce the VQC rotation angles and the observable parameters, respectively.

\section{Experiments}
\begin{figure}[htbp]
\begin{center}
\includegraphics[width=1\columnwidth]{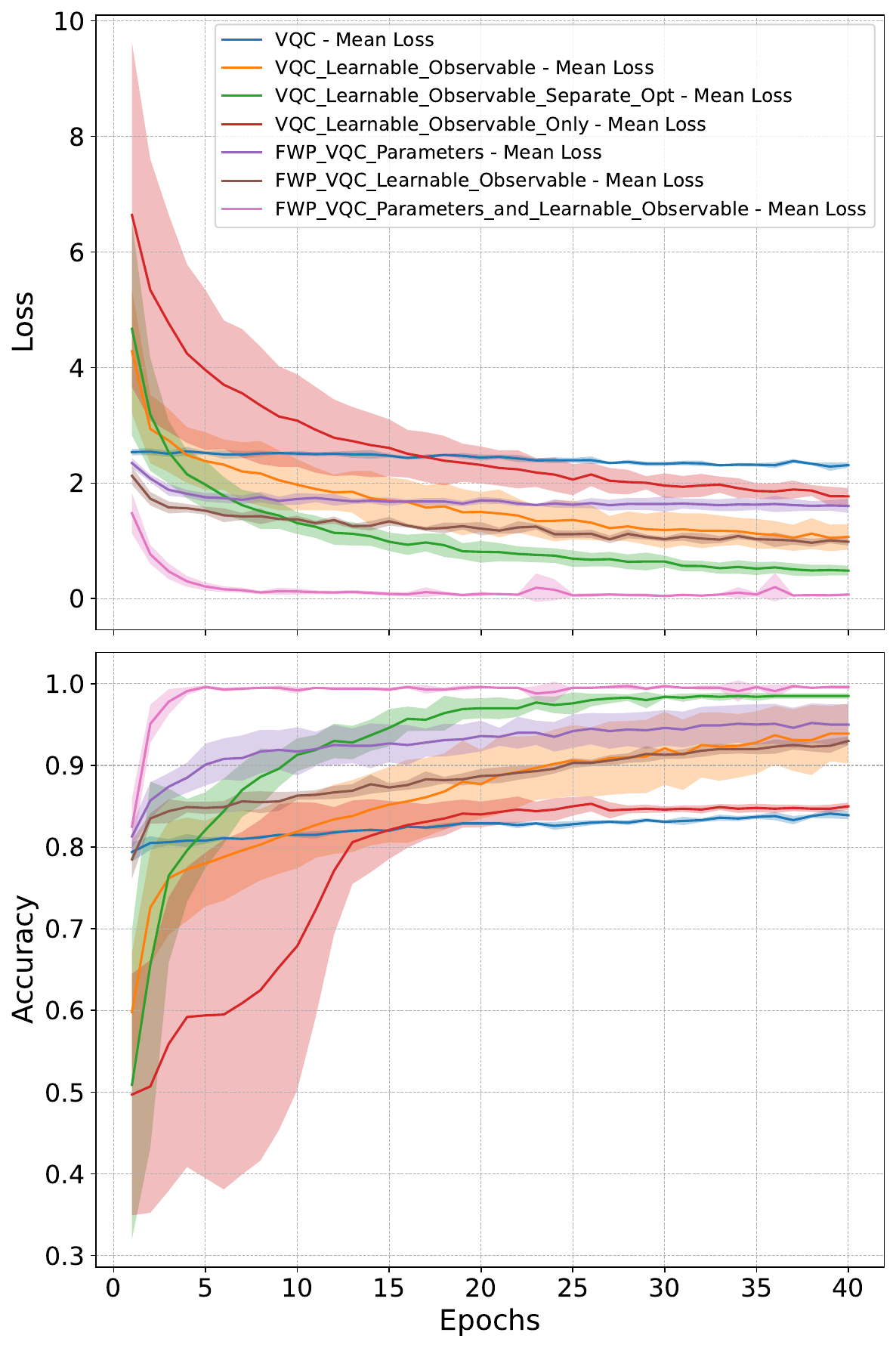}
\caption{{\bfseries Comparison of different VQC models in \texttt{make\_moons} dataset with noise = 0.1.} Top: Mean training loss across 5 random seeds (± standard deviation). Bottom: Mean classification accuracy (± standard deviation). FWP-based configurations, particularly \emph{FWP VQC Parameters and Learnable Observable}, achieve faster convergence and higher accuracy, suggesting that neural generation of both circuit parameters and observables enhances training efficiency and robustness.}
\label{fig:results_make_moons_noise_0.1}
\end{center}
\end{figure}

To evaluate the effectiveness of the proposed method, we conduct a series of experiments under various VQC learning configurations, as summarized in \tableautorefname{\ref{tab:baselines_and_architectures}}. In the table, if the ``Observable'' field is marked as ``None'', a fixed Pauli-$Z$ operator is used as the measurement observable. If the ``Circuit Param'' is labeled ``None'', the corresponding VQC parameters are randomly initialized and remain untrained throughout. Entries marked as ``NN'' indicate that the corresponding component is dynamically generated by an external neural network controller. All experiments follow a unified training protocol unless otherwise specified: batch size = 20, training set size = 200, testing set size = 100, VQC depth = 2. The learning rate is set to 0.01 for all FWP-based methods and for direct optimization of circuit parameters, while a higher learning rate of 0.1 is used for optimizing observables directly. Each experiment is repeated five times with different random seeds to report the mean and standard deviation for robust statistical comparison. For the \texttt{make\_moons} and \texttt{make\_circles} tasks, we employ a 4-qubit system. In the \texttt{make\_classification} experiments, the number of qubits is matched to the input feature dimension, using 8, 10, and 12 qubits, respectively.
\begin{table}[htbp]
\caption{\bfseries{Various VQC learning settings considered.} }
\label{tab:baselines_and_architectures}
\resizebox{\columnwidth}{!}{%
\begin{tabular}{|l|l|l|}
\hline
                                            & Circuit Param & Observable \\ \hline
VQC                                         & RMSProp       & None       \\ \hline
VQC Learnable Observable                    & RMSProp       & RMSProp    \\ \hline
VQC Learnable Observable Separate Opt       & RMSProp       & Adam       \\ \hline
VQC Learnable Observable Only               & None          & RMSProp    \\ \hline
FWP VQC Parameters                          & NN            & None       \\ \hline
FWP VQC Learnable Observable                & None          & NN         \\ \hline
FWP VQC Paramaters and Learnable Observable & NN            & NN         \\ \hline
\end{tabular}%
}
\end{table}
\begin{figure}[htbp]
\begin{center}
\includegraphics[width=1\columnwidth]{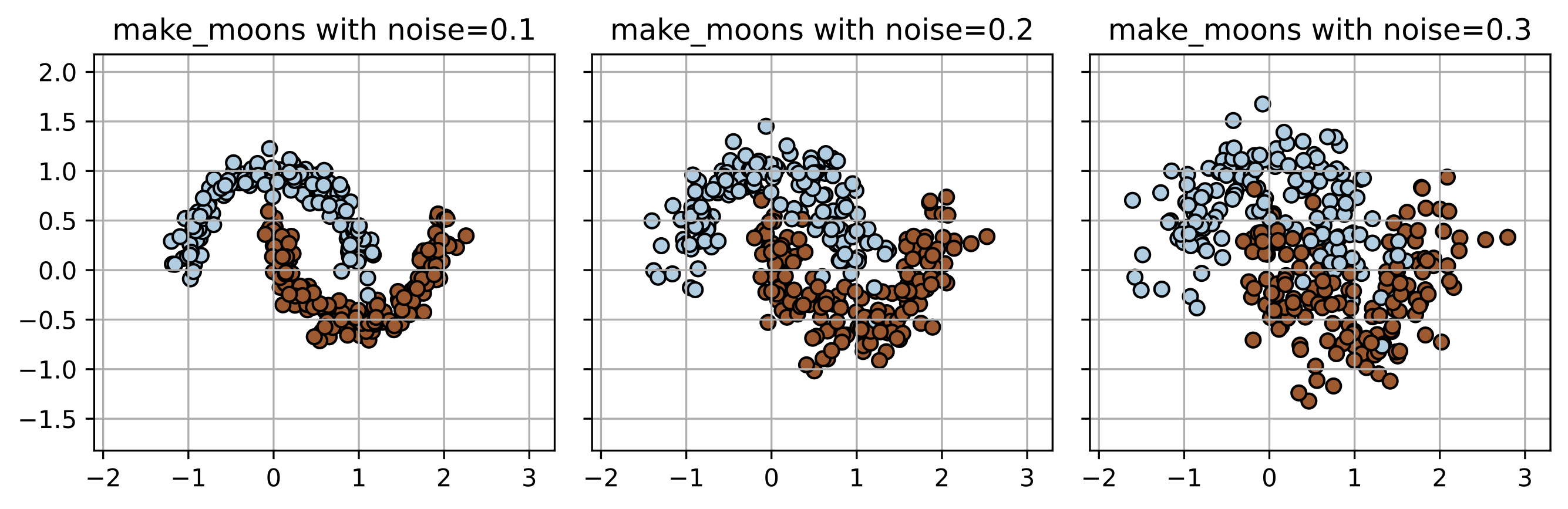}
\caption{{\bfseries Visualization of  \texttt{make\_moons} dataset with different noise configurations.}}
\label{fig:make_moons_visualization}
\end{center}
\end{figure}
To evaluate the consistency and effectiveness of various VQC learning strategies, we conduct a series of experiments on the \texttt{make\_moons} dataset with Gaussian noise ($\text{noise} = 0.1$). Each configuration is trained across five different random seeds, and the reported results in \figureautorefname{\ref{fig:results_make_moons_noise_0.1}} show the mean and standard deviation of both training loss and classification accuracy over 40 epochs. The results reveal a clear advantage for FWP-based models over traditional optimizer-based VQCs. Notably, the setting where both the circuit parameters and measurement observable are generated by a data-conditioned neural network (\emph{FWP VQC Parameters and Learnable Observable}) achieves the most rapid convergence and highest accuracy, while maintaining low variance across seeds. In contrast, traditional VQCs that rely solely on classical optimizers, especially those optimizing only the observable (\emph{VQC Learnable Observable Only}), exhibit slower and less stable learning. These results highlight the benefits of one-shot parameter generation and further support the hypothesis that decoupling parameter optimization from iterative procedures can mitigate gradient-related training challenges in quantum models.

\begin{figure}[htbp]
\begin{center}
\includegraphics[width=1\columnwidth]{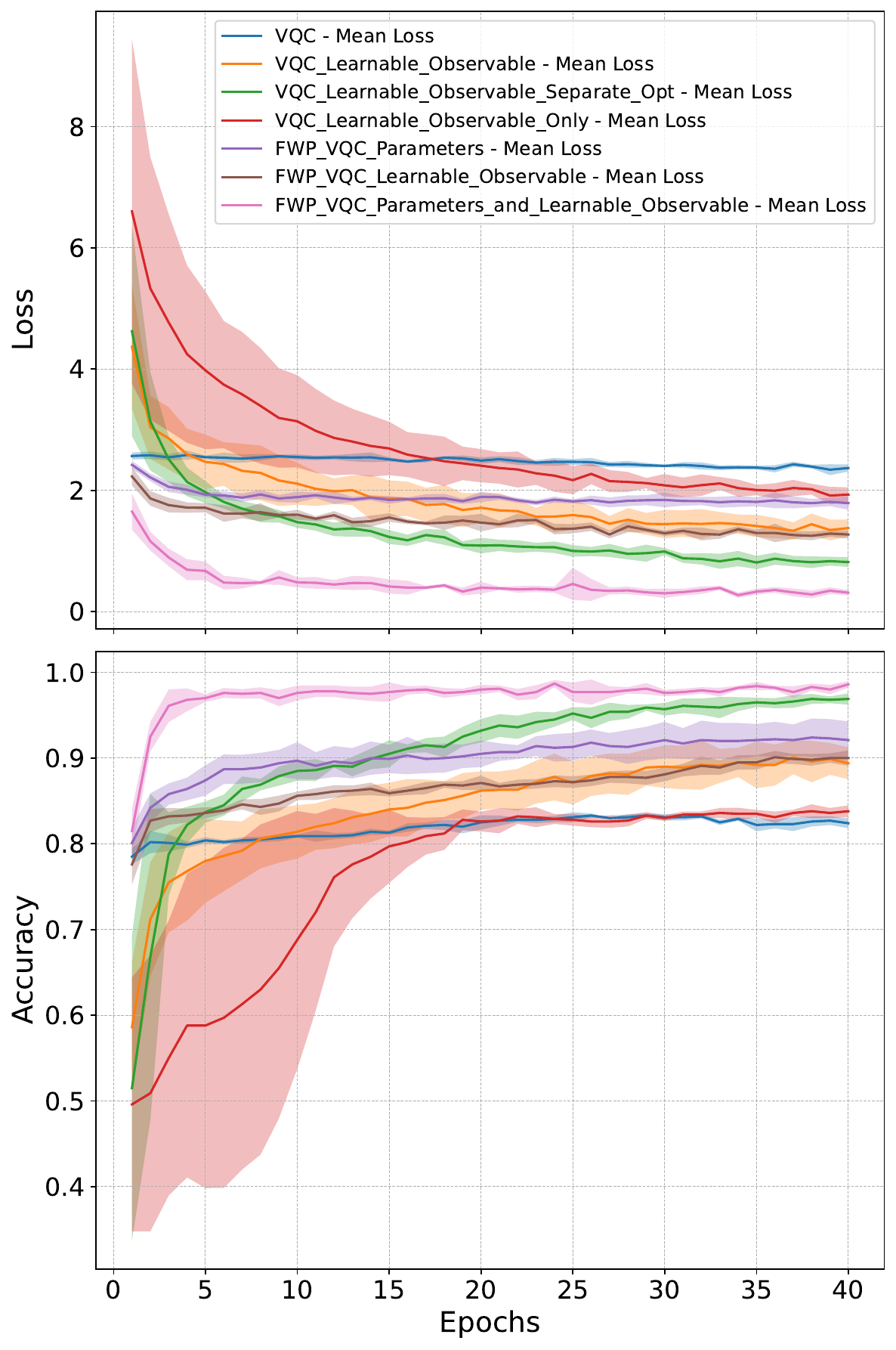}
\caption{{\bfseries Comparison of different VQC models in \texttt{make\_moons} dataset with noise = 0.2.} Top: Mean training loss over 5 runs (± standard deviation). Bottom: Classification accuracy across 40 epochs. The \emph{FWP VQC Parameters and Learnable Observable} setting maintains superior performance despite the added noise, achieving near-perfect accuracy with low variance.}
\label{fig:results_make_moons_noise_0.2}
\end{center}
\end{figure}

To evaluate model robustness under increased input uncertainty, we repeat the classification experiment with a higher noise level ($\text{noise} = 0.2$) in the \texttt{make\_moons} dataset. As shown in \figureautorefname{\ref{fig:results_make_moons_noise_0.2}}, all models experience performance degradation to varying degrees, but the proposed FWP-based configurations remain substantially more resilient. In particular, \emph{FWP VQC Parameters and Learnable Observable} maintains a rapid convergence rate and achieves near-perfect accuracy across runs. The low variance bands indicate consistent performance across different seeds, even in the presence of noisy inputs. This suggests that one-shot parameter generation conditioned on the input distribution may offer a regularizing effect and improve generalization. Traditional VQCs optimized via classical optimizers suffer more from the increased noise, with the baseline \emph{VQC} and \emph{VQC Learnable Observable Only} showing higher variance and slower accuracy growth. These results reinforce the benefits of integrating neural modules (FWP) to bypass direct optimization of quantum parameters, particularly in scenarios with data uncertainty.

\begin{figure}[htbp]
\begin{center}
\includegraphics[width=1\columnwidth]{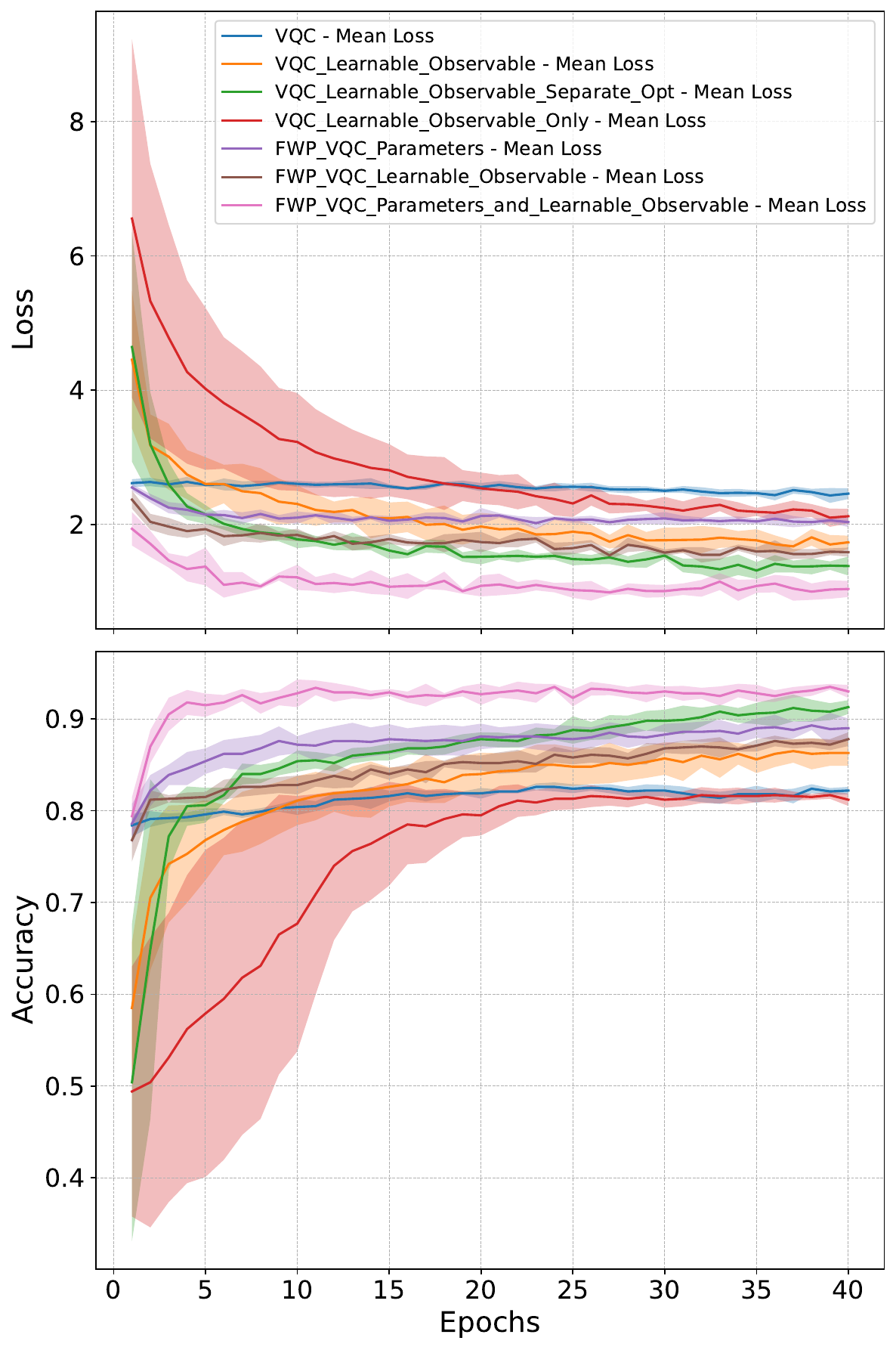}
\caption{{\bfseries Comparison of different VQC models in \texttt{make\_moons} dataset with noise = 0.3.} Top: Mean training loss over 5 random seeds ($\pm$ standard deviation). Bottom: Mean classification accuracy ($\pm$ standard deviation). Despite increased data noise, FWP-based models, especially the dual-generation variant, maintain strong accuracy and low loss variance, demonstrating superior robustness compared to traditional optimizer-based VQCs.}
\label{fig:results_make_moons_noise_0.3}
\end{center}
\end{figure}

\begin{figure}[htbp]
\begin{center}
\includegraphics[width=1\columnwidth]{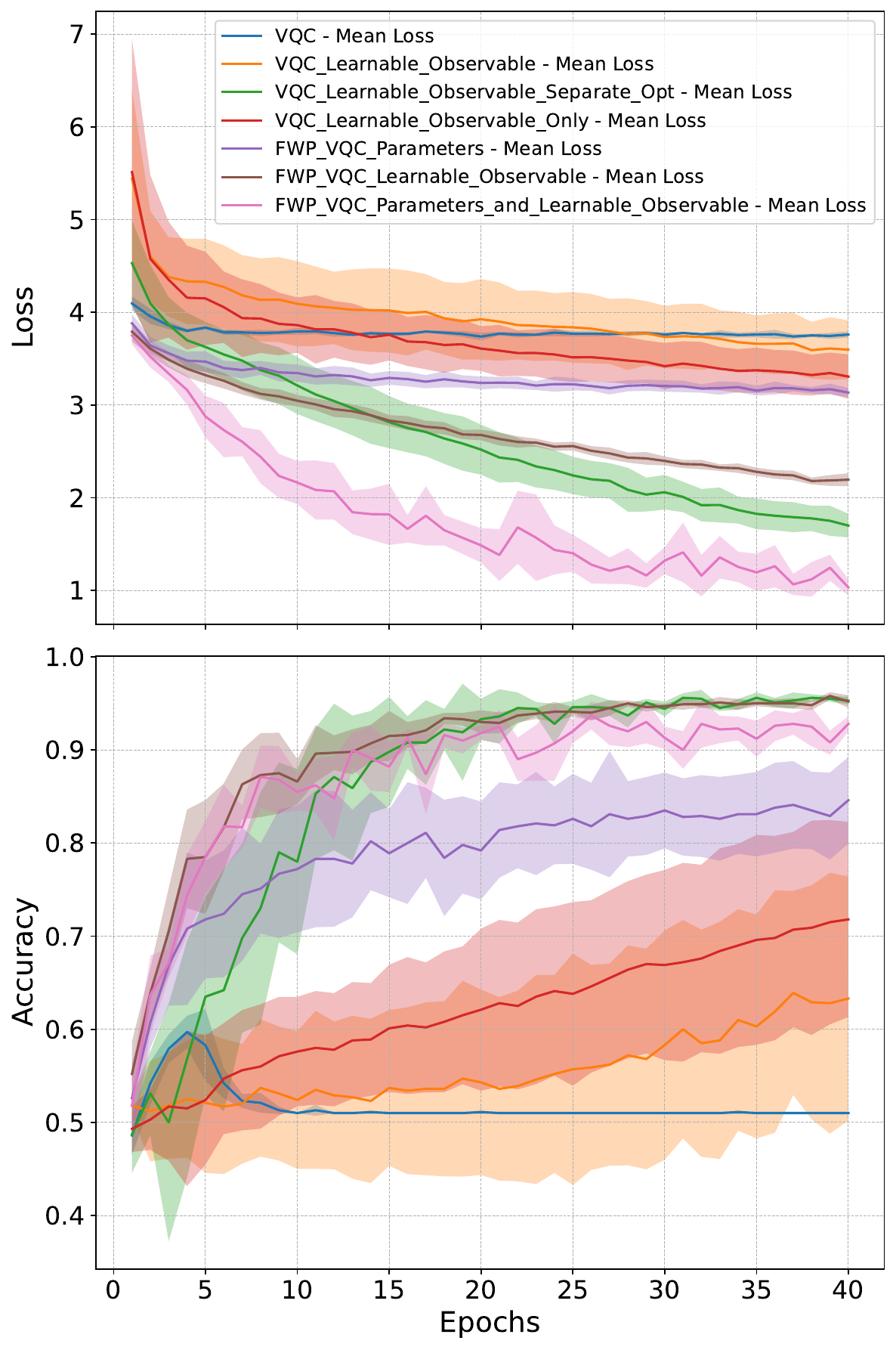}
\caption{{\bfseries Comparison of different VQC models in \texttt{make\_circles} dataset with noise = 0.05.} Top: Mean training loss over 5 random seeds ($\pm$ standard deviation). Bottom: Mean classification accuracy. While FWP-based models achieve faster convergence, the \emph{VQC Learnable Observable Separate Opt} setting also reaches competitive final accuracy, suggesting that careful optimizer assignment can partially mitigate classical limitations.}
\label{fig:results_make_circles_noise_0.05}
\end{center}
\end{figure}

\begin{figure}[htbp]
\begin{center}
\includegraphics[width=1\columnwidth]{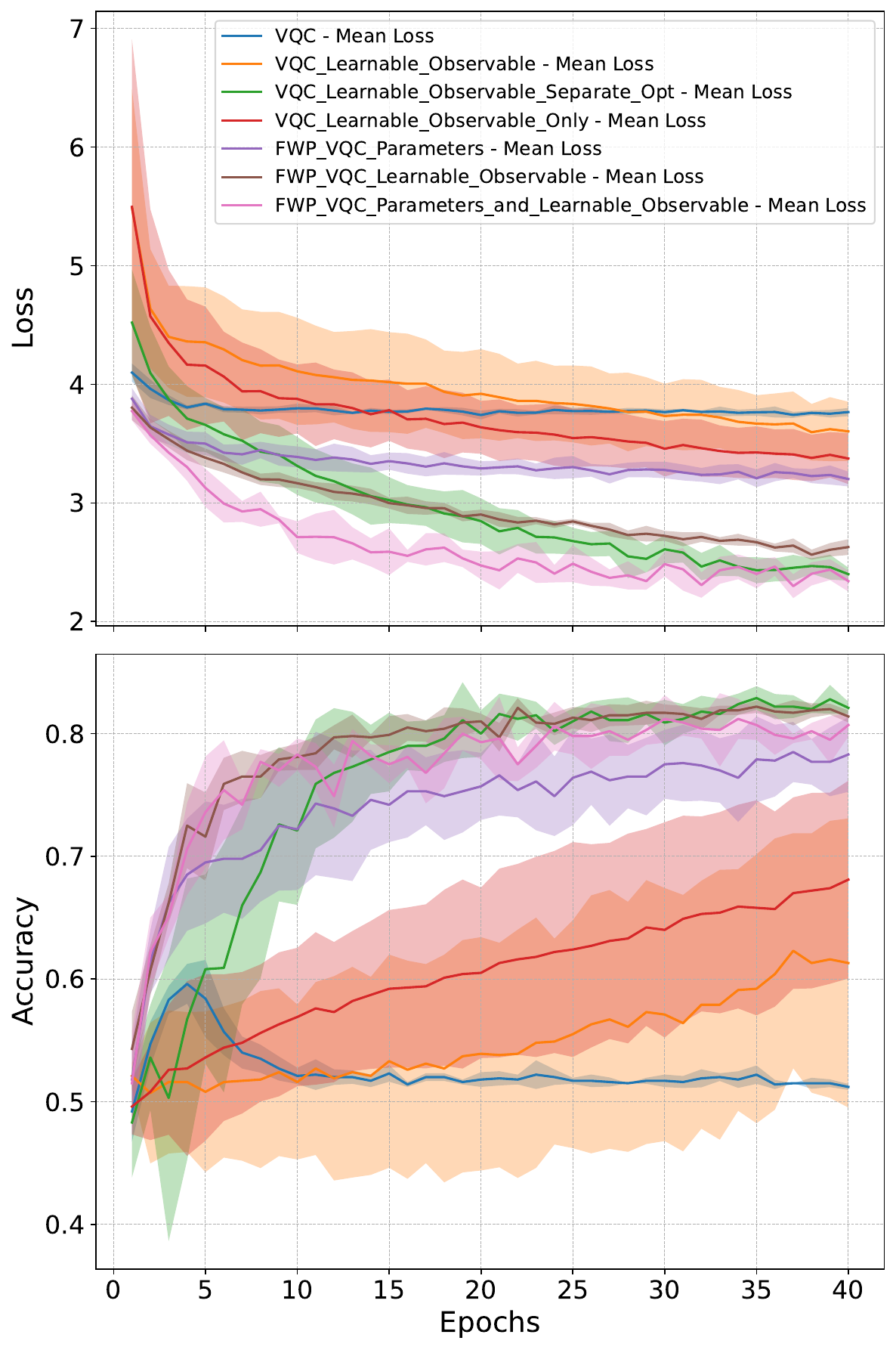}
\caption{{\bfseries Comparison of different VQC models in \texttt{make\_circles} dataset with noise = 0.1.} Top: Mean loss over 5 random seeds ($\pm$ standard deviation). Bottom: Accuracy progression over 40 epochs. Classical optimizer-based VQCs exhibit reduced stability, while FWP-based models retain both faster convergence and superior accuracy. Notably, the dual-generator model consistently outperforms all baselines.}
\label{fig:results_make_circles_noise_0.1}
\end{center}
\end{figure}

\begin{figure}[htbp]
\begin{center}
\includegraphics[width=1\columnwidth]{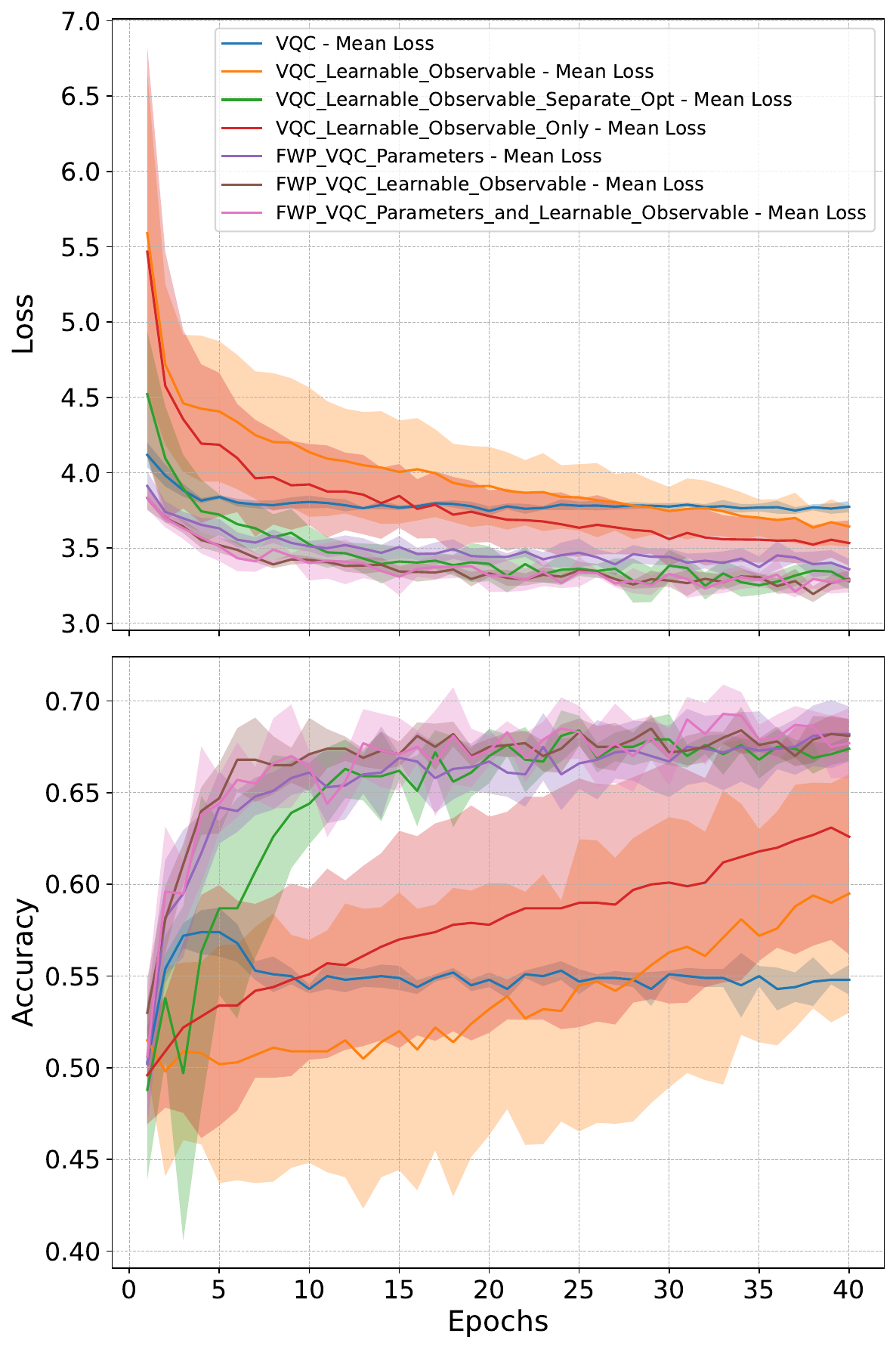}
\caption{{\bfseries Comparison of different VQC models in \texttt{make\_circles} dataset with noise = 0.2.} Top: Mean training loss (± std) across 5 runs. Bottom: Classification accuracy over 40 epochs. With increased noise, classical optimizer-based VQCs exhibit performance collapse, while FWP-based models retain relatively stable accuracy and learning dynamics.}
\label{fig:results_make_circles_noise_0.2}
\end{center}
\end{figure}

\begin{figure}[htbp]
\begin{center}
\includegraphics[width=1\columnwidth]{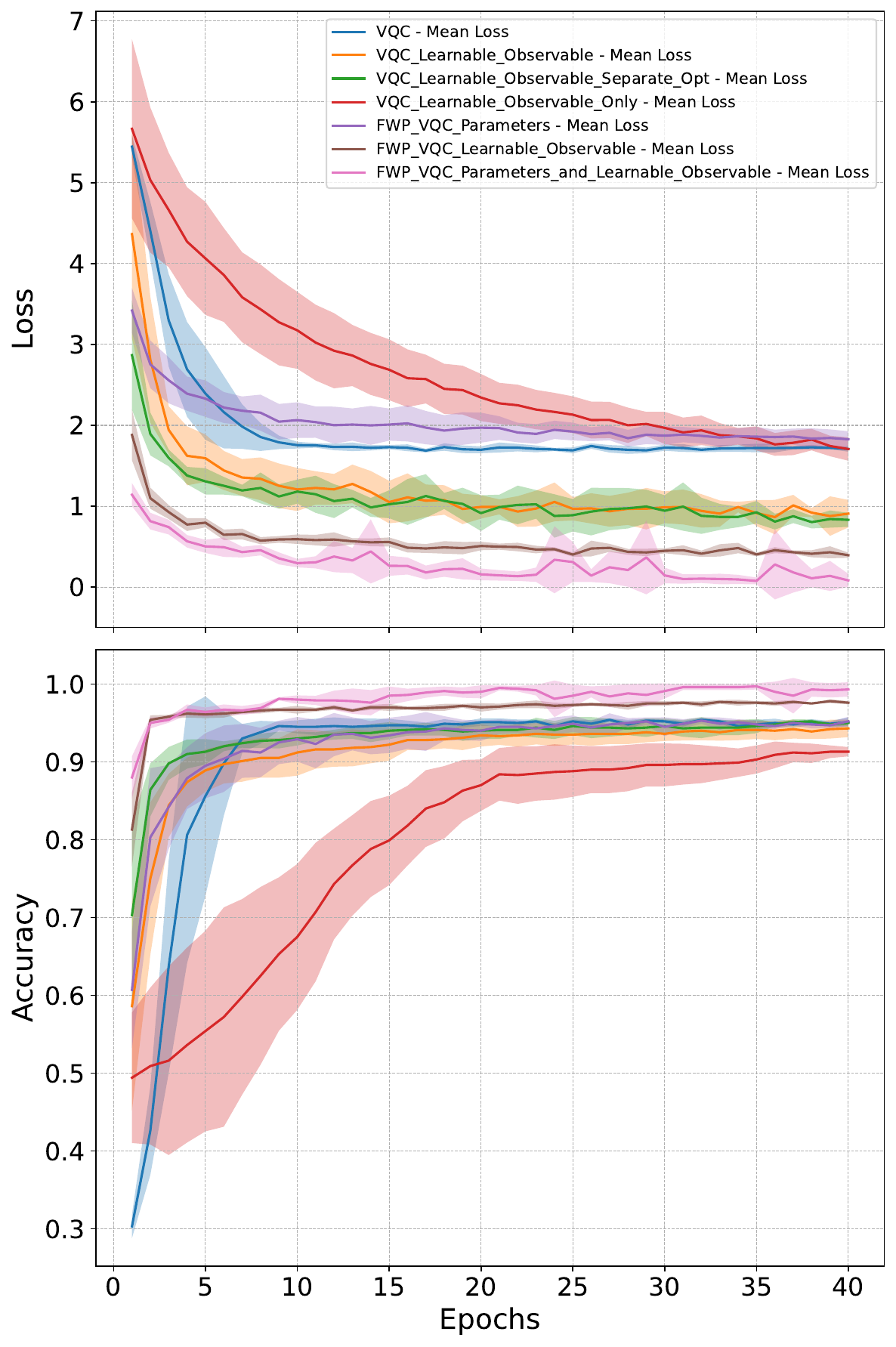}
\caption{{\bfseries Comparison of different VQC models in \texttt{make\_classification} dataset with 8 features.} Top: Mean training loss over 5 random seeds ($\pm$ standard deviation). Bottom: Classification accuracy. All models eventually reach high accuracy, but FWP-based configurations—especially the dual-generator-achieve significantly faster convergence and lower loss throughout training.}
\label{fig:results_make_classification_8q}
\end{center}
\end{figure}

\begin{figure}[htbp]
\begin{center}
\includegraphics[width=1\columnwidth]{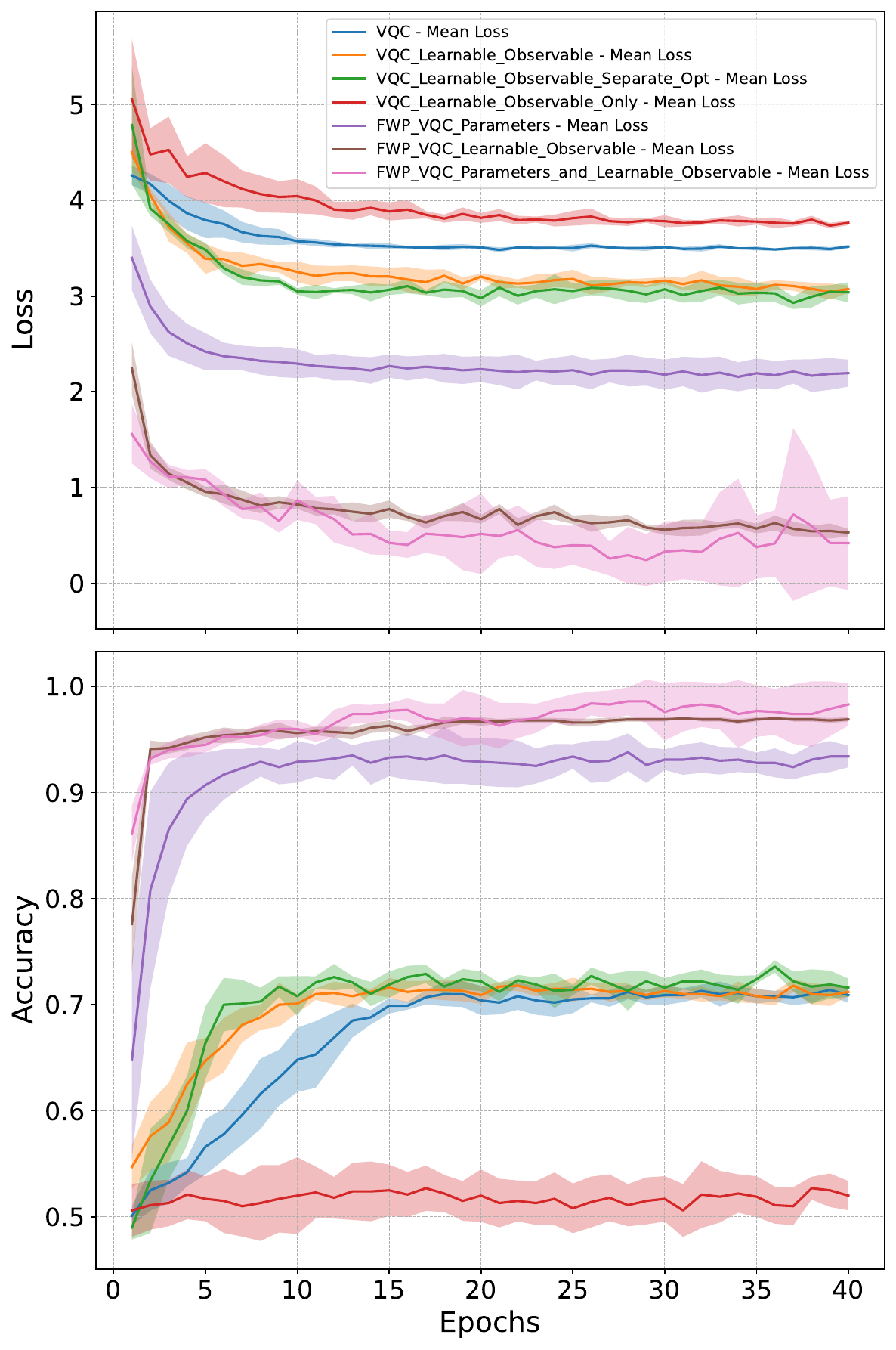}
\caption{{\bfseries Comparison of different VQC models in \texttt{make\_classification} dataset with 10 features.} Top: Mean loss over 5 random seeds (± standard deviation). Bottom: Classification accuracy over training epochs. As input dimensionality increases, traditional optimizer-based models plateau below 75\% accuracy, while FWP-based configurations continue to scale effectively and reach over 95\%.}
\label{fig:results_make_classification_10q}
\end{center}
\end{figure}

\begin{figure}[htbp]
\begin{center}
\includegraphics[width=1\columnwidth]{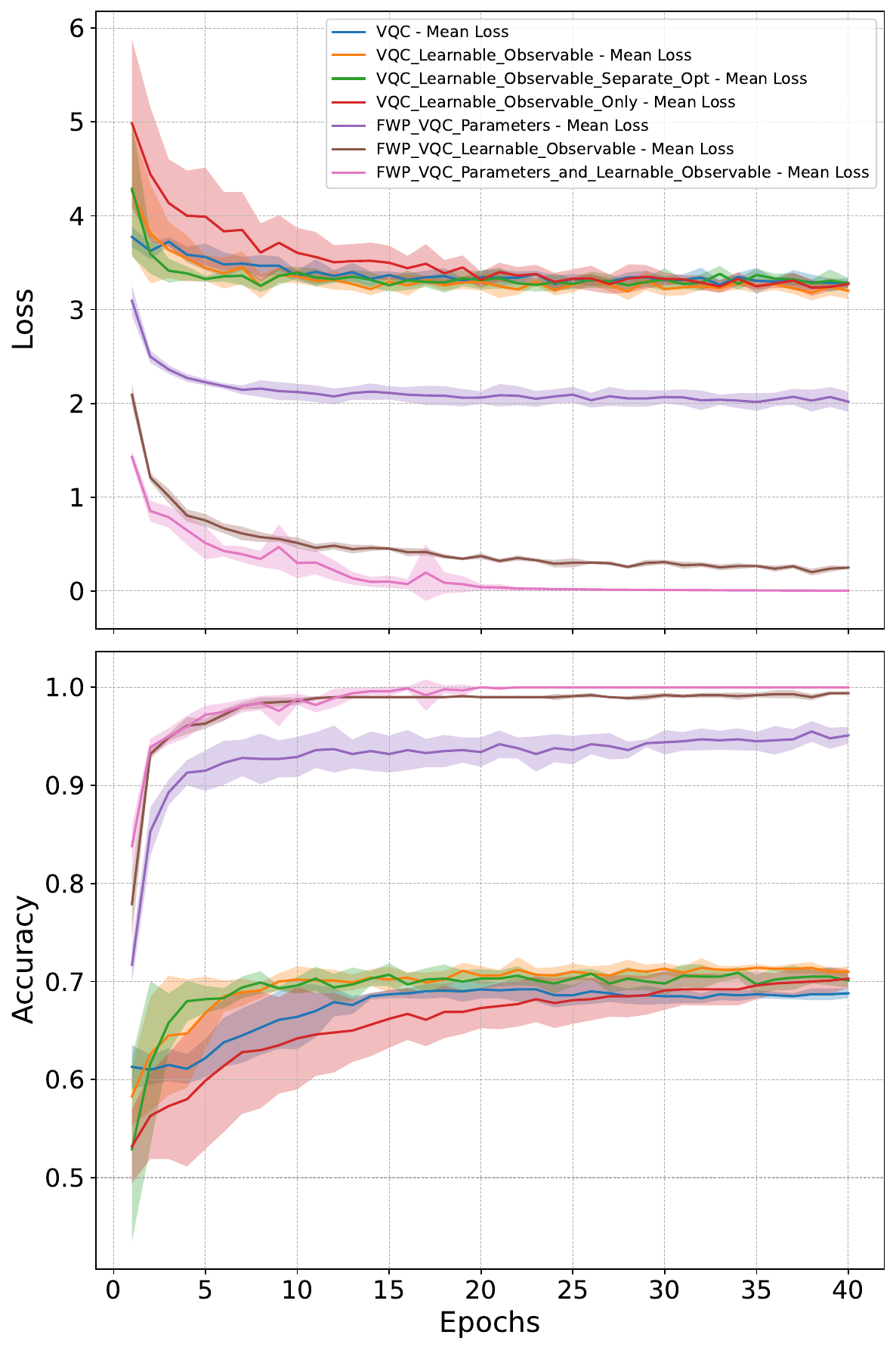}
\caption{{\bfseries Comparison of different VQC models in \texttt{make\_classification} dataset with 12 features.} Top: Mean loss across 5 seeds (± standard deviation). Bottom: Classification accuracy. Classical optimizer-based VQCs fail to scale effectively, while all FWP-based models maintain prediction performance, with the dual-generator achieving near-perfect accuracy.}
\label{fig:results_make_classification_12q}
\end{center}
\end{figure}

To further examine model robustness under challenging conditions, we repeat the classification task with $\text{noise} = 0.3$, simulating severe input uncertainty. As shown in \figureautorefname{\ref{fig:results_make_moons_noise_0.3}}, traditional optimization-based VQCs exhibit noticeable degradation, with slower convergence and wider performance variance. Notably, the \emph{VQC Learnable Observable Only} configuration becomes increasingly unstable, suggesting that training observables in isolation may be insufficient in high-noise regimes. By contrast, FWP-based architectures remain highly resilient. The configuration utilizing neural generation for both circuit parameters and observables (\emph{FWP VQC Parameters and Learnable Observable}) continues to outperform other settings, achieving near 0.95 accuracy with minimal variance across all seeds. These results highlight the potential of neural-controlled quantum models to generalize well even under adverse conditions.
Across all noise levels ($\text{noise} = 0.1$, $0.2$, and $0.3$), the proposed FWP-based configurations consistently outperform their classical counterparts. In particular, the full FWP model that generates both VQC rotation parameters and observables maintains high accuracy, low loss, and reduced variance regardless of data perturbations. These findings suggest that integrating a data-conditioned neural prior into VQC training may enhance generalization and stability in real-world noisy quantum learning scenarios.
\begin{figure}[htbp]
\begin{center}
\includegraphics[width=1\columnwidth]{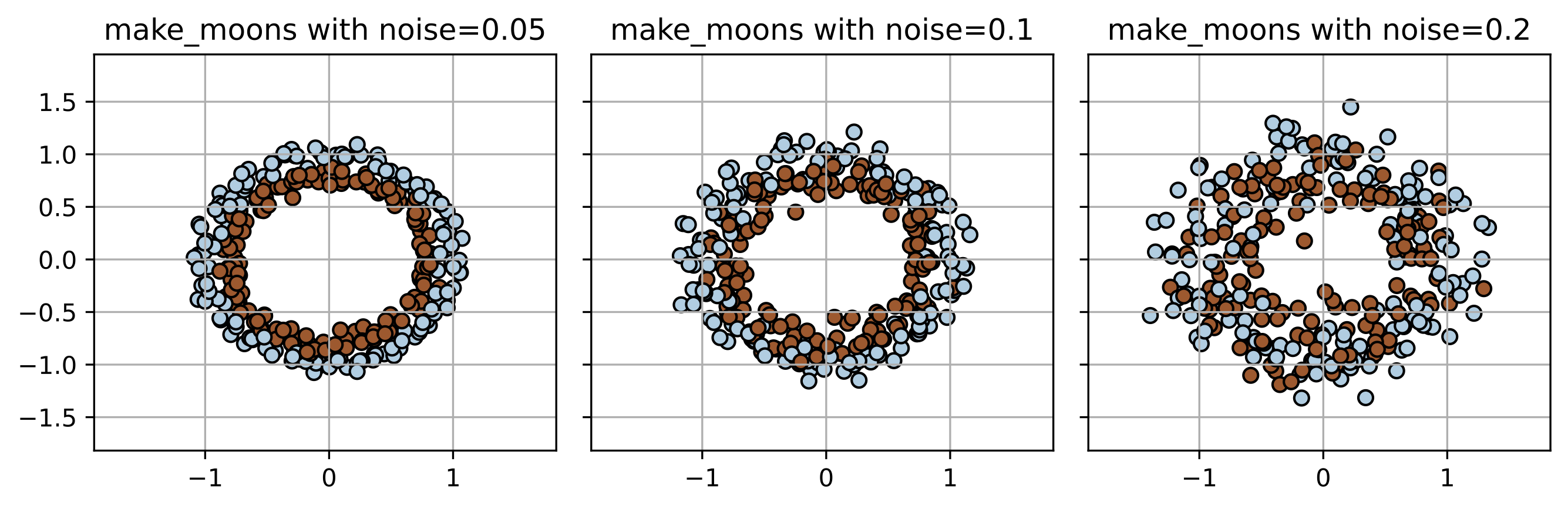}
\caption{{\bfseries Visualization of  \texttt{make\_circles} dataset with different noise configurations.}}
\label{fig:make_circles_visualization}
\end{center}
\end{figure}
On the \texttt{make\_circles} dataset with low noise ($\text{noise} = 0.05$), the benefit of FWP-based configurations remains evident in terms of rapid convergence and early-stage accuracy (shown in \figureautorefname{\ref{fig:results_make_circles_noise_0.05}}). Notably, \emph{FWP VQC Parameters and Learnable Observable} achieves the fastest loss reduction and reaches high accuracy within the first few epochs. However, unlike previous experiments, the \emph{VQC Learnable Observable Separate Opt} setting performs surprisingly well in this task, ultimately matching or slightly exceeding the final accuracy of FWP models. This indicates that separating the optimization strategies for circuit parameters and observables may lead to better convergence behavior in classically trained VQCs-at least under low-noise, geometrically structured inputs. These results suggest that while neural parameter generation (FWP) offers clear advantages in generalization and early-stage learning, certain hand-crafted optimizer configurations can still compete under favorable conditions.
When the noise level is increased to $\text{noise} = 0.1$ in the \emph{make\_circles} dataset, the gap between FWP-based configurations and classical VQCs becomes more evident (\figureautorefname{\ref{fig:results_make_circles_noise_0.1}}). While the \emph{VQC Learnable Observable Separate Opt} setting still reaches competitive final accuracy, its early-stage learning is slower and its performance variance is significantly higher. In contrast, the FWP configurations—especially the dual-generator (\emph{FWP VQC Parameters and Learnable Observable})-achieve rapid convergence, stable training curves, and the highest overall accuracy. Even under moderate noise, classical baselines such as \emph{VQC} and \emph{VQC Learnable Observable} suffer from underfitting and instability. These findings highlight the robustness of data-conditioned quantum parameter generation under noisy conditions, and suggest that FWP-based architectures can serve as more reliable foundations in real-world applications where clean data is not guaranteed.
When evaluated under high noise conditions ($\text{noise} = 0.2$) in the \texttt{make\_circles} dataset, most classical VQC settings exhibit significant performance degradation (\figureautorefname{\ref{fig:results_make_circles_noise_0.2}}). The standard \emph{VQC} and \emph{VQC Learnable Observable Only} configurations fail to generalize and produce accuracy curves barely exceeding 55\%. While \emph{VQC Learnable Observable Separate Opt} had previously shown promise, its performance becomes increasingly unstable in this noisy regime. In contrast, the FWP-based models—especially the dual-generator setting—maintain stable training curves and achieve the highest accuracy among all tested configurations, with minimal variance. These results reinforce the hypothesis that neural parameter generation mechanisms such as FWP are more resilient to noisy inputs, offering a promising direction for quantum machine learning in real-world, imperfect data environments.
In the synthetic \texttt{make\_classification} task with 8 input features, all models are able to reach high final accuracy, reflecting the relatively low task complexity. However, significant differences arise in training dynamics and convergence speed (\figureautorefname{\ref{fig:results_make_classification_8q}}). The \emph{FWP VQC Parameters and Learnable Observable} configuration achieves near-perfect accuracy within the first few epochs, with the lowest average loss and smallest variance. Other FWP configurations also converge rapidly, outperforming classical optimizer-based settings in both early-stage and stable performance. In contrast, traditionally optimized VQCs-especially \emph{VQC} and \emph{VQC Learnable Observable}-exhibit slower convergence and higher loss curves, highlighting the inefficiency of classical gradient methods in encoding meaningful representations early in training. The \emph{VQC Learnable Observable Only} configuration continues to underperform, validating the need for full-model adaptability in hybrid quantum-classical setups.
In the more challenging setting with 10 input features from the \texttt{make\_classification} dataset, the performance gap between FWP-based models and classical VQC configurations becomes increasingly pronounced (\figureautorefname{\ref{fig:results_make_classification_10q}}). While the \emph{VQC Learnable Observable Separate Opt} previously showed competitive accuracy, it now plateaus near 70\%, indicating its limited scalability to higher-dimensional feature spaces. Conversely, all FWP-based models maintain strong generalization and fast convergence, with \emph{FWP VQC Parameters and Learnable Observable} achieving nearly perfect classification performance. Although this dual-generator model shows minor fluctuations in loss, its accuracy remains consistently superior. The \emph{FWP VQC Parameters} variant also demonstrates robust training stability, striking a favorable balance between efficiency and reliability. These results affirm the scalability of FWP-based approaches to more complex learning tasks, highlighting their potential for practical applications involving high-dimensional quantum-classical data.
As the number of input features increases to 12, classical VQC models exhibit further degradation in both convergence and final accuracy (\figureautorefname{\ref{fig:results_make_classification_12q}}). Most traditional settings, including \emph{VQC}, \emph{VQC Learnable Observable}, and \emph{VQC Learnable Observable Separate Opt}, plateau at ~70\% accuracy and struggle to reduce training loss. In contrast, FWP-based models maintain their performance advantage and demonstrate excellent scalability. The \emph{FWP VQC Parameters and Learnable Observable} configuration once again achieves near-perfect accuracy and minimal training variance. The \emph{FWP VQC Learnable Observable} setting also performs exceptionally well, followed closely by \emph{FWP VQC Parameters}, which remains above 90\% accuracy throughout training. These findings highlight the structural limitation of classical optimizer-based approaches when facing increased input complexity, and further validate the effectiveness of FWP-based architectures in high-dimensional quantum learning.
\section{Conclusion}
In this study, we conducted a comprehensive ablation analysis of various learning configurations for variational quantum circuits (VQCs), with a particular focus on classifying synthetic datasets under different noise levels and input dimensionalities. Our comparisons span classical optimizer-based approaches and neural network-based parameter generation, implemented via the proposed Fast Weight Programmer (FWP). Numerical results across multiple benchmark tasks—including \texttt{make\_moons}, \texttt{make\_circles}, and \texttt{make\_classification}-consistently demonstrate the superiority of FWP-based configurations. The dual-generator variant, which jointly produces VQC parameters and observables conditioned on input data, achieves rapid convergence, low variance, and superior classification accuracy even under high noise or feature-rich settings. In contrast, classical VQC training pipelines struggle with scalability, often failing to generalize in noisy or complex regimes. These findings suggest that replacing classical optimizers with input-aware neural generation mechanisms offers a promising direction for training hybrid quantum-classical models more efficiently. The proposed FWP framework not only alleviates gradient-based optimization bottlenecks but also improves robustness and adaptability in quantum learning tasks.
\bibliographystyle{IEEEtran}
\bibliography{bib/qml_examples,bib/vqc,bib/qas,bib/qt,bib/fwp,bib/qc,bib/classical_ml,bib/learnable_measurement}
\end{document}